\documentclass[pre,aps,twocolumn]{revtex4}
\usepackage{amsfonts,amssymb,amsmath,bm}
\usepackage[dvips]{graphicx}
\usepackage{color}
\usepackage[unicode=true]{hyperref}
\usepackage{times}
\usepackage{siunitx}

\begin{document}

\title{Acoustic streaming in 2D freely suspended smectic liquid crystal film}

\author{S.V. Yablonskii$^{1,2}$, N.M. Kurbatov$^{2,3}$, V.M. Parfenyev$^1$}
\email{parfenius@gmail.com}

\affiliation{$^1$Landau Institute for Theoretical Physics of the Russian Academy of Sciences, 142432 Akademica Semenova av. 1-A, Chernogolovka, Russia; \\
$^2$Shubnikov Institute of Crystallography of Federal Scientific Research Center "Crystallography and Photonics" of the Russian Academy of Sciences, 119333 Leninsky pr. 59, Moscow, Russia; \\ $^3$Moscow Institute of Physics and Technology, 141700 Institutskiy per. 9, Dolgoprudny, Russia.}

\date{\today}

\begin{abstract}
We study a horizontal streaming excited by means of a low frequency and intensity acoustic wave in 2D freely suspended film of thermotropic smectic liquid crystal. Acoustic pressure induces fast periodic transverse oscillations of the film, which produce in-plane stationary couples of vortices slowly rotating in opposite directions owing to hydrodynamic nonlinearity. The parameters of vortices are measured using a new method, based on tracking disk-shaped island defects. The horizontal motion occurs only when the amplitude of acoustic pressure exceeds threshold value, which can be explained by Bingham-like behavior of the smectic film. The measurements above threshold are in a good agreement with existing theoretical predictions. We have demonstrated experimentally that in-plane flow is well controlled by changing the acoustic pressure, excitation frequency and geometry of the film. The observations open a way to use the phenomenon in various applications.
\end{abstract}

\maketitle

\textit{Introduction.} --- Discovered by Otto Lehmann and Friedrich Reinitzer more than 100 years ago, thermotropic liquid crystals are used everywhere \cite{Vorlander}. Lyotropic liquid crystals, substances similar to ordinary soap combined with water, were known much earlier and their importance in everyday life is also not in doubt \cite{Huck}.
The main application of thermotropic liquid crystals is the production of displays \cite{slide}. Other applications (so-called "non-display" field) include, e.g., the development of panels with controllable light transmission \cite{craighead1982new} and visualization of thermal fields in medicine and aeronautics \cite{Hallcrest}. Lyotropic liquid crystals are used as drug and as means of drug delivery \cite{chen2014cubic}, in the form of grease with ultra-low friction \cite{pogodina14}, and to simulate the processes occurring in the natural cell membranes \cite{blinov88}.
Thermotropic smectic liquid crystals possess microscopic layered structure that promotes to the formation of macroscopic steady state in the form of uniform in thickness freely suspended film without contact with bearing surfaces \cite{pieranski93}. The bending motion of such films are well-studied and its properties allow to use smectic films as sensitive elements in detectors of visible and IR radiation \cite{detector}.

In this paper, we investigate acoustically stimulated horizontal streaming in freely suspended smectic films \cite{parfenyev16} and discuss possible applications of the phenomenon.
The earliest example of acoustic streaming refers to vortex spots in soap films \cite{taylor78}, and it is close to experiments discussed below. An important and fundamental difference between the acoustic streaming in thermotropic smectic films and in soap films is the lack of tangential stresses associated with the Marangoni effect \cite{vega98} in the first case. Here, acoustic streaming originated in the constant thickness membrane composed of an individual mesogenic substances.
The in-plane motion have a form of stationary vortices rotating in opposite directions, which can be controlled by parameters of acoustic wave and geometry of the film. The acoustic wave excites transverse oscillations of the film, which involve surrounding air in motion. In the linear approximation, the air motion has horizontal vorticity arising due to viscosity and concentrated in a thin viscous sublayer near the film surface. As a second order nonlinear effect the horizontal vorticity is slightly tilted due to surface tilt of the film, and this mechanism produces a vertical vorticity --- the vortices in the film plane. Note that similar mechanism generates horizontal vortices on a surface of liquid \cite{16FPVBLL}.

The vortex flow is analyzed by observing the motion of disk-shaped island defects generated on the film surface in a special way \cite{pattanaporkratana04}. We demonstrate that the horizontal streaming occurs only if the amplitude of acoustic pressure exceeds a certain value. The vortex motion arises abruptly, the velocity of defects has a discontinuity at threshold. Above threshold the amplitude of streaming is proportional to the squared acoustic pressure in accordance with existing theory \cite{parfenyev16}. We discuss the threshold behaviour and speculate that it can be explained by Bingham-like mechanism \cite{bingam}.
Experimental data also shows that 2D freely suspended smectic films are sufficiently robust and can withstand moderate levels of mechanical shock without degradation of their characteristics. This allows to use them in various applications, which are discussed later in the paper.

\begin{figure}[t!]
\center{\includegraphics[width=\linewidth]{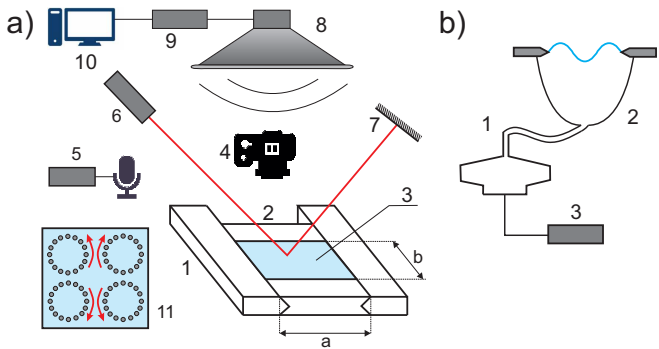}}
\caption{a) Experimental setup for the study of vibration and vortex motion in the liquid crystal membrane: 1, guide rails; 2, mobile barriers; 3, liquid crystal membrane; 4, digital camera; 5, condenser microphone connected to acoustic power meter; 6, laser; 7, screen; 8, loudspeaker; 9, low-frequency generator based on PC sound card; 10, computer; 11, vortices visualized by island defects; b) Acoustic excitation by the waveguide: 1, sound waveguide; 2, frame with the membrane; 3, sound generator.}
\label{fig:1}
\end{figure}

\textit{Methods.} --- Smectic freely suspended film was prepared by manipulation with mobile barriers \cite{pieranski93} as shown in Fig.~\ref{fig:1}a (1-3). A drop of liquid crystal was placed in a narrow gap between two barriers, which then were slowly moved to a predetermined size. During experiments, the barriers can be shifted relatively to each other changing the aspect ratio of rectangular film. The distance between guide rails had been fixed at $a = 16.5$ mm and the distance between mobile barriers $b$ was ranged from $0$ to $60$ mm. The thickness of the freely suspended smectic film was determined by the color in reflected white light, as was done by Sirota \textit{et al.} \cite{sirota87}. The film color was blue, corresponding to the number of layers $N = 62$ and film thickness $h = Nd \approx 198$ nm, where each smectic layer has thickness $d = 3.2$ nm \cite{oswald15}. By selecting a certain initial amount of the liquid crystal we always managed to obtain a film of uniform thickness with acceptable accuracy. Additional confirmation of the film thickness was performed by fitting the theoretical reflection spectrum and the reflection spectrum obtained by the CCD AvaSpec-ULS2048 spectrometer \cite{pieranski93}. The measurements were performed with the liquid crystal n-octyl-cyano-biphenyl 8CB, which exhibits $S_A$ phase between $21.0$ and $33.0^{\circ}C$. The purity of 8CB (NIOPIK) was $99\%$. The working temperatures were $22-28^{\circ}C$.

Transverse vibrations of the smectic film and the vortex flow were excited by an electrodynamic loudspeaker ($8$) acting as a piston, see Fig.~\ref{fig:1}a. Frame with the film was placed in the far field, which was uniform with an accuracy of $3 \%$ on the acoustic pressure amplitude. The frequency and amplitude of the acoustic wave were controlled by computer ($10$). The acoustic pressure was measured by condenser microphone connected to the sound level meter Robotron-$017$ ($5$). In the experiments we varied the size of membrane $b$, the excitation frequency $\nu$ and the amplitude of acoustic pressure $\Delta p$. The thickness of the film with acceptable accuracy was constant.
We also investigated the horizontal streaming excited by strongly inhomogeneous acoustic field formed by means of an acoustic waveguide, as shown in Fig.~\ref{fig:1}b. In this case, the amplitude of acoustic heterogeneity on the surface of membrane reached $25\%$.

To study the horizontal flow in the smectic film we tried various known methods of adding passive particles on the film surface. Monitoring the motion of these particles it is possible to restore the velocity field in the film plane \cite{parfenyev16_2}.
Attempts were made to use the micron-sized polystyrene beads, but the film was either destroyed or particles were collected on the periphery of the film in the region of meniscus. The known method of creating colored liquid for monitoring flow using photochromic labels \cite{popovich67} was also ineffective, because the film thickness is too small and therefore one cannot reach the necessary contrast.
Eventually, we found a convenient way to visualize the flow by observing the motion of island defects generated on the film surface, see Fig.~\ref{fig:1}a ($11$). To generate these defects mobile barriers ($2$) were abruptly shrunk by about $1-2$ mm. The procedure allows to get controllably different amount of long-lived defects, whose sizes are in the range of $0.01$ to $0.5$ mm.

\begin{figure}[t]
\begin{center}
\includegraphics[width=0.95\linewidth]{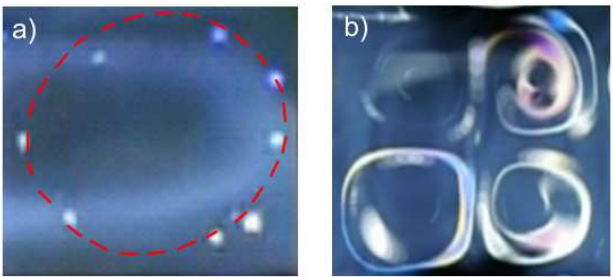}
\caption{Horizontal streaming at high acoustic pressure: a) The vortex visualized by island defects. The dotted line shows trajectory of defects; b) Four vortices visualised by an excess amount of defects.}
\label{fig:3}
\end{center}
\end{figure}

At high acoustic pressure (more than $\sim 0.16$ Pa for used parameters, see below) these defects cannot be treated as passive tracers. For example, they were gathered together on the same current line regardless of their formation place, see Fig.~\ref{fig:3}a and \cite[1]{supplement}. Whereas at lower acoustic pressure, they manifest Lagrangian particle behavior quite well, and their motion provides information about the in-plane flow in the bed of standard theory. The transition between two regimes is shown in video \cite[2]{supplement}.
The dignity of proposed observation method can be seen with the naked eye, compare Fig.~\ref{fig:3}a and Fig.~\ref{fig:3}b, \cite[1]{supplement} and \cite[3]{supplement}. The last ones show a vortex motion in the case of visualization by a large number of defects and high acoustic pressure. Similar in quality pictures were observed earlier in experiments with soap films \cite{soap}.

\textit{Results.} --- Stationary vortex flow is generated due to nonlinear interaction of bending modes excited in the film \cite{parfenyev16}. Therefore, the acoustic streaming occurs near the frequencies corresponding to the resonance conditions of smectic film transverse oscillations:
\begin{eqnarray}\label{eq:1}
&\nu_{nm} = \dfrac{|k_{nm}|}{2 \pi} \sqrt{\dfrac{2 \sigma}{\rho_s}}, \quad \rho_s = \rho_{LC} h + 2 \rho_a / |k_{nm}|, &
\end{eqnarray}
where $|k_{nm}| = \pi \sqrt{(n/a)^2+(m/b)^2}$ is wavenumber of the bending mode, $\sigma = 29.2$ dyn/cm is the surface tension of the liquid crystal 8CB, $\rho_s$ is a surface density of the film including the mass of air involved in motion. Volume densities of liquid crystal 8CB and air are equal to $\rho_{LC} = 0.96$ g/cm$^3$ and $\rho_a(25^{\circ}C) = 1.2 \cdot 10^{-3}$ g/cm$^{3}$ respectively. The plane acoustic wave effectively excites only modes $(n,m)$ with odd indices \cite{pieranski93}.
The essential point in the experimental observation of coherent vortex structures is small "detuning" in the geometry of the film, it should be nearly square to produce a time-shift between excited bending modes \cite{parfenyev16}. Further, we consider such film, $a=16.5$ mm and $b=17$ mm, where bending modes $(3,1)$ and $(1,3)$ are excited. Using Eq.~(\ref{eq:1}) one finds resonance frequencies of these modes, $\nu_{31} = 357$ Hz and $\nu_{13} = 344$ Hz. Note that the mass of air involved in motion is much larger than the mass of the smectic film, i.e. $2 \rho_a/\rho_{LC}hǀkǀ \approx 20$.

\begin{figure*}[t!]
\begin{center}
\includegraphics[width=\linewidth]{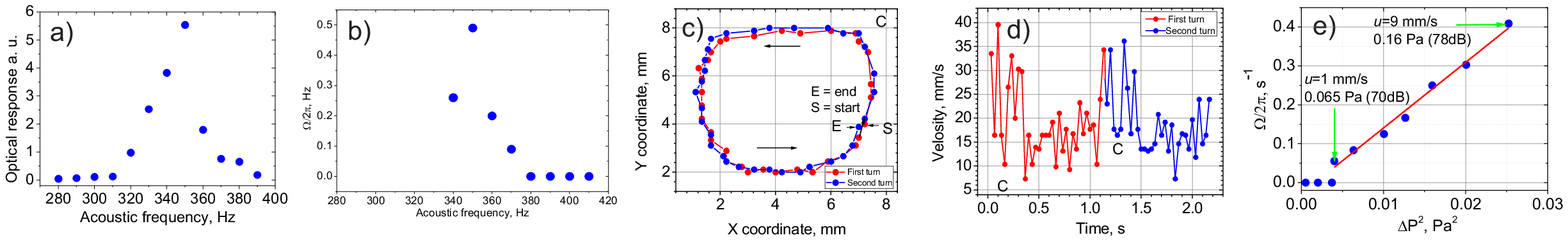}
\caption{Experimental results of the acoustic streaming study for slightly rectangular film, $a=16.5$ mm, $b= 17$ mm: a) The size of laser beam projection on a screen after reflection from a surface of the film vs. the frequency of acoustic excitation; b) Angular velocity of defects depending on the acoustic frequency, $\Delta p=0.2$ Pa; c) The trajectory of the island defect, point C is the membrane center, $\Delta p=0.2$ Pa, $\nu = 350$ Hz; d) Velocity of island defect on the trajectory depending on time, point C corresponds to passage of defect near the membrane center, $\Delta p=0.2$ Pa, $\nu = 350$ Hz; e) Island defect angular velocity depending on the squared amplitude of the acoustic pressure, $\nu=350$ Hz.}
\label{fig:2}
\end{center}
\end{figure*}

Experimentally, resonance frequency close to $350$ Hz was recorded, see Fig.~\ref{fig:2}a. It was obtained by measuring the size of laser beam projection on a screen after reflection from a surface of the film vs. frequency of the acoustic excitation, see Fig.~\ref{fig:1}a (6-7). The $Q$-factors of resonances $(3,1)$ and $(1,3)$ are not very high and they merge into a single curve. The indices of excited modes were verified by examining optical images of laser beam reflected from different parts of the membrane as explained in \cite[4]{supplement}. It is important to emphasise that during the measurements of optical responses the vortex motion was absent.

The vortices appear only when the amplitude of acoustic pressure reaches a certain value. In-plane vortex flow was visualized by island defects generated on the film surface, as was explained earlier and shown in Fig.~\ref{fig:3}a. Analyzing the motion of defects captured by a digital camera we were able to calculate the average angular velocity $\Omega$ of tracing particles.
Fig.~\ref{fig:2}b presents the frequency dependence of this velocity $\Omega$ near the resonance frequency $\nu = 350$ Hz. Before disappearing defects had enough time to make about several tens of revolutions. It is clearly seen that frequency dependence of the angular velocity $\Omega$ correlates quite well with the resonance curve of bending modes, see Fig.~\ref{fig:2}a. This demonstrates that the vortex motion is generated due to transverse oscillations of the smectic film.

Fig.~\ref{fig:2}c and Fig.~\ref{fig:2}d show the typical trajectory of defects and their velocity on the trajectory respectively at acoustic pressure higher than $\sim 0.16$ Pa, when the tracers are not Lagrangian particles as was explained earlier. The trajectory of particles was slightly asymmetrical and differs from a circle because of interaction between vortex structures. Each point on the figures was obtained as a result of time averaging over about $10$ film oscillations. The time interval between two consecutive shots of CCD camera was $1/30$ sec, while the period of membrane oscillations was $1/350$ sec.
Fig.~\ref{fig:2}d demonstrates the complex motion of defects on a regular trajectory, the membrane center (point C) serves as a stop point, where defects slow down significantly.

Fig.~\ref{fig:2}e demonstrates the threshold behavior of the defect's velocity vs. amplitude of acoustic pressure for some particular defect. For the indicated pressures defects are distributed on the film surface uniformly, see \cite[2]{supplement}, and each defect has unique trajectory. We believe that in this case the tracers can be approximately treated as passive particles. The dependence $\Omega(\Delta p)$ has a pronounced yield stress $\Delta p_0 = 0.063$ Pa, which slightly depends on concentration of island defects. Above threshold the tracked defect abruptly begins to move with some finite velocity. This fact reflects the non-Newtonian behavior of smectic crystals belonging to a class of Bingham liquids \cite{bingam}. In Bingham liquid when a shear stress is below some threshold value, only elastic behavior is observed. Above threshold one observes an onset of plastic flow. It is well known that in liquid crystal the viscoplastic behavior occurs when it contains elastic lattice, composed of interacted topological defects, e.g., coupled edge dislocations \cite{bingam, fujii11}. Consequently, the flow is structurally suppressed and the acoustic pressure below threshold provokes only finite deformations of the film.
It should be underlined that in our experiment the motion of island defects appears abruptly, whereas in Bingham body the flow occurs gradually. Evidently, such behavior corresponds to the abrupt destruction of the lattice of topological defects.

\textit{Discussion.} --- The theory of 2D vortex motion in acoustically excited freely suspended smectic films has been developed in the paper \cite{parfenyev16}. The key point of this approach was consideration of two time-shifted degenerate bending modes, excited in the film and coupled by nonlinear interaction. There was obtained an explicit expression for the vertical vorticity $\varpi$ on the film surface in terms of the film displacement. The main result of theoretical approach can be written in the form:
\begin{equation}\label{eq:3}
  \varpi = A (x, y) \Delta p^2 \sin \phi,
\end{equation}
where the phase shift $\phi$ describes time-shift between bending modes and $A(x,y)$ characterises the vorticity spatial distribution. The restoration of vertical vorticity by observing the motion of passive particles on the film surface was discussed in the paper \cite{parfenyev16_2}. It turns out that the measured vorticity can be described by the same Eq.~(\ref{eq:3}) up to numerical factor.

Experimental results are in a good agreement with theoretical predictions. For excited bending modes $(3,1)$ and $(1,3)$ the theory predicts four counter-rotating vortices and we have observed them experimentally, see Fig.~\ref{fig:3}b. Next, according to the theory the phase shift $\phi$ should be equal to zero for the square film and there is no vorticity in that case. Changing the aspect ratio of the frame near that point, $a/b = 1$, we have observed that vortices changed the direction of their rotation as was predicted, see \cite[5]{supplement}. We have also showed that above threshold the angular velocity $\Omega$ of the defect is proportional to the squared acoustic pressure $\Delta p$ in accordance with expression (\ref{eq:3}), see Fig.~\ref{fig:2}e. Thus, we have proved experimentally that vortex motion at the film surface is a second order nonlinear effect.

Highly inhomogeneous acoustic excitation provides even more powerful horizontal streaming control. It allows to change spatial distribution of vortices and obtain more intense currents. We propose to use the phenomenon of acoustic streaming for mixing small amounts of matter. The rotational speed of such mixer reaches $600$ rev/min and the direction of rotation can be tuned by acoustic frequency. Modulation of the acoustic excitation signal can be used as an additional control parameter. Note that under the same conditions we observed coalescence effect, which can find applications as particles manipulation technique. See supplemental videos \cite[6-8]{supplement} for illustration.
We also would like to add that we were unable to get a stable vortex pattern in the case of too narrow frame, e.g. with a width $b < 5$ mm, see also \cite{brazovskaya96}. The reason for the suppression of vortex motion in the case of small size frames probably is the strong influence of the solid boundary including the meniscus named in \cite{qi16} as ''wall'' effect. Most of the work remains to be done to explain the phenomenon in detail.



It is also necessary to say a few words about two-dimensionality of freely suspended films. Soap films are not strictly 2D because their in-plane flow is accompanied by a change in film thickness \cite{danilov00}. Nevertheless, these films were considered as model systems providing an opportunity to test 2D turbulence theory. The enstrophy cascade from large scales to small scales was confirmed experimentally on soap films \cite{kellay98}. However, the detailed analysis has shown that the fluid motion in soap films is more complex and that a relation to motion described by the 2D hydrodynamic equations is not straightforward \cite{couder89}. On the other hand, the smectic thermotropic films, retaining a constant thickness during the flow, are free from these shortcomings. Let us consider slow motion excited in the smectic film. From the incompressibility condition we obtain that the velocity $\bm u$ is directed in $X-Y$ film plane, i.e. $u_z \sim u kh \ll u$, where $h$ is a film thickness and $1/k$ is characteristic scale of horizontal motion, e.g. the size of vortex. Based on experimental measurements, we can also estimate the variation of horizontal velocity $\delta u$ with a film thickness. For highly inhomogeneous acoustic excitation we obtain $\eta \delta u/h \sim \Delta p$, where $\eta = 5 \cdot 10^{-2}$ Pa$\cdot$s is smectic dynamic viscosity \cite{viscosity} and $\Delta p \sim 0.5$ Pa is acoustic pressure, and thus $\delta u \sim 2 \cdot 10^{-3}$ mm/s. At the same parameters, the measured horizontal velocity was $u \sim 25$ mm/s, and therefore the velocity field in thin smectic films is two-dimensional with a high accuracy.

In conclusion, we have presented a first-ever experimental observation of an acoustic streaming in the smectic liquid crystal films. 2D hydrodynamic flow in the form of vortex couples was studied by observing the motion of island defects generated on the film surface. The threshold character of vortex motion was found. We associate threshold of acoustic streaming with a Bingham-like behavior of the liquid crystal membrane, but the phenomenon requires further investigation. We believe that smectic freely suspended films can be used to study the relationship between the microscopic properties of lamellar materials and their viscoelastic behavior. In addition, we propose the implementation of smectic freely suspended films for the study of 2D hydrodynamics and in the models of large global circulation \cite{Seychelles}. Our findings can be also used for providing driving power for microfluidic mixers.

\acknowledgments

We are grateful to V. Lebedev, E. Kats, and S. Vergeles for valuable discussions.
This work was supported by the Russian Science Foundation (project no. 14-12-00475).

\end{document}